\documentclass[%
 reprint,
superscriptaddress,
 amsmath,amssymb,
 aps,
]{revtex4-1}

\usepackage{graphicx}
\usepackage{dcolumn}
\usepackage{bm}
\usepackage{xcolor}

\newcommand*\BLogo{\includegraphics[width=0.75\baselineskip]{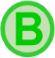}}
\newcommand*\PLogo{\includegraphics[width=0.75\baselineskip]{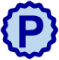}}

\begin{document}

\preprint{APS/123-QED}

\title{Spontaneous Pulse Formation in Edge-Less Photonic Crystal Resonators}

\author{Su-Peng Yu}
 \email{supeng.yu@colorado.edu}
 \affiliation{Time and Frequency Division, NIST, Boulder, CO 80305, USA}
 \affiliation{Department of Physics, University of Colorado, Boulder, CO 80309, USA}

\author{Daniel C. Cole}
 \affiliation{Time and Frequency Division, NIST, Boulder, CO 80305, USA}
  \affiliation{Department of Physics, University of Colorado, Boulder, CO 80309, USA}

\author{Hojoong Jung}
 \affiliation{Time and Frequency Division, NIST, Boulder, CO 80305, USA}
  \affiliation{Department of Physics, University of Colorado, Boulder, CO 80309, USA}
 
 \author{Gregory T. Moille}
\affiliation{Microsystems and Nanotechnology Division, NIST, Gaithersburg, MD 20899, USA}
 \affiliation{Joint Quantum Institute, NIST/University of Maryland, College Park, MD 20742, USA}

  \author{Kartik Srinivasan}
 \affiliation{Microsystems and Nanotechnology Division, NIST, Gaithersburg, MD 20899, USA}
 \affiliation{Joint Quantum Institute, NIST/University of Maryland, College Park, MD 20742, USA}
 
 \author{Scott B. Papp}
 \affiliation{Time and Frequency Division, NIST, Boulder, CO 80305, USA}
  \affiliation{Department of Physics, University of Colorado, Boulder, CO 80309, USA}

\date{\today}

\begin{abstract}
Complex systems are a proving ground for fundamental interactions between components and their collective emergent phenomena. Through intricate design, integrated photonics offers intriguing nonlinear interactions that create new patterns of light. In particular, the canonical Kerr-nonlinear resonator becomes unstable with a sufficiently intense traveling-wave excitation, yielding instead a Turing pattern composed of a few interfering waves. These resonators also support the localized soliton pulse as a separate nonlinear stationary state. Kerr solitons are remarkably versatile for applications, but they cannot emerge from constant excitation. Here, we explore an edge-less photonic-crystal resonator (PhCR) that enables spontaneous formation of a soliton pulse in place of the Turing pattern. We design a PhCR in the regime of single-azimuthal-mode engineering to re-balance Kerr-nonlinear frequency shifts in favor of the soliton state, commensurate with how group-velocity dispersion balances nonlinearity. Our experiments establish PhCR solitons as mode-locked pulses by way of ultraprecise optical-frequency measurements, and we characterize their fundamental properties. Our work shows that sub-wavelength nanophotonic design expands the palette for nonlinear engineering of light.

\end{abstract}

\maketitle

Integrated nonlinear photonics is a versatile engine to generate and control electromagnetic radiation, opening new application directions and enabling fundamental studies. Second- and higher-order nonlinear susceptibilities now form the basis of many photonics technologies; a good example is harmonic-\cite{Hickstein2017} or difference-frequency\cite{Tadanagaa2006} generation that realize laser sources from the ultraviolet to the infrared. In particular,  third-order, Kerr processes are ubiquitous in photonics due to intensity dependence of the refractive index, $n=n_0 + n_2\,I$, where $n_2$ is the nonlinear index and $I$ is intensity. They enable spontaneous formation of stationary configurations of electromagnetic fields that affect conversion of a laser from one color to another. More generally, modulation instability that arises from nonlinearity governs interesting behaviors in systems ranging from quantum matter \cite{Carr2004} to desert sand dunes \cite{Parteli2011}. Studying nonlinear behaviors in the exquisitely controlled environment of integrated photonics-- where sub-wavelength features lead to new optical behaviors --can increase understanding in a variety of physical systems.

Kerr resonators-- optical cavities built from an $n_2$ material --are an attractive system for fundamental studies and applications. We understand the formation of some pattern and pulse states of the intraresonator field $\psi$ from the Lugiato-Lefever equation (LLE)  \(\partial_\tau \psi = -(1+i\alpha)\psi - \frac{i}{2}\beta \partial_\theta^2 \psi +   i|\psi|^2\psi +F\), where $\theta$ is the resonator angular coordinate, $- \frac{i}{2}\beta \partial_\theta^2 \psi$ is the group-velocity dispersion (hereafter GVD or dispersion), $|\psi|^2 \psi$ is the nonlinearity, $F$ is a traveling-wave pump-laser field originating outside the resonator with a red detuning by $\alpha$ to lower frequency than the resonator mode; see Ref. \cite{Godey2014} for further details. A few states stand out amongst the diverse solution space of the LLE \cite{Godey2014}: The constant-amplitude flat state energized by a sufficiently weak pump laser; the Turing pattern that emerges when the flat state is unstable; and the Kerr soliton that is a localized pulse coexisting with, but not emerging spontaneously from, the flat state. Indeed, microresonator soliton frequency combs \cite{Kippenberg2018} have been engineered to support a wide range of applications, including optical communication\cite{MarinPalomo2017, Fulop2018}, spectroscopy \cite{Suh2016}, and ranging \cite{Trocha2018}. GVD engineering via the cross-sectional waveguide dimensions offers powerful control of soliton properties \cite{Yu2019FP}. Moreover, exotic photonic states have been reported using unconventional resonator-mode engineering \cite{Lobanov2015, Xue2015}. 

Spontaneous formation of patterns from break-up of the flat state is a critical outcome in the LLE. A pattern forms spontaneously by four-wave mixing (FWM), constrained by a balance of the Kerr frequency shift $\delta_\mu$ of the comb mode number $\mu$, and the phase-mismatch from dispersion $\frac{1}{2} \beta \, \mu^2$. We count the comb modes and the resonator modes with respect to the mode closest to the pump laser (hereafter the pump mode, $\mu=0$). Importantly, $\delta_\mu$ for each mode depends on the intraresonator field according to \( \delta_\mu = g\, (2\,N- |a_\mu|^2 )\) \cite{FactorOfTwoBook}, where $a_\mu$ are the Fourier decomposition amplitude for mode $\mu$, $g$ the per-photon Kerr shift, and $N$ the total photon number. The term $g=1$ is a standard normalization of the LLE. Beginning with the flat state, all $a_{\mu'\neq 0}=0$ and $\delta_{\mu=0} = 2 N - N = \frac{1}{2} \delta_{\mu'\neq 0}$, where the modes $\mu'$ are not pumped. The difference between self- and cross-phase modulation results in a reduced Kerr shift for the pump mode by a factor of two compared to other modes. This reduced Kerr shift enables FWM for the Turing pattern at modes $\pm\mu'$, characterized by \(\frac{1}{2} \beta |\mu'|^2 - \delta_{\pm\mu'} = - \delta_{\mu=0} \). Conversely, the soliton is a collective state with many modes $\mu'$ that reach phase-matching only at large $\alpha$ where the flat-state amplitude is insufficient to support spontaneous FWM processes. These phase-matching conditions result in the disparate generation behaviors of Turing patterns and solitons.


Here, we explore a re-balancing of the LLE that causes Kerr-soliton formation from break-up of the flat state, replacing the Turing pattern. To accomplish this dramatic outcome, we design and fabricate edge-less photonic-crystal resonators (PhCR), which are Kerr-microresonators with their inner wall modified by an azimuthally uniform, nanopatterned shape oscillation. \textcolor{black}{The ring geometry imposes the edge-less boundary condition on the photonic waveguide, opening the PhCR bandgap -- thus controllably shifting the frequency -- for one azimuthal mode.} We program the shift to directly phase-match the soliton with the pump laser nearly on-resonance of the pump mode. Moreover, this shifts the Turing pattern off-resonance, precluding its formation. We have realized and explored spontaneous soliton formation in wide-ranging experiments, including observing the immediate transition from the flat state to the soliton, soliton pulse bandwidth control by dispersion engineering through the bulk ring dimensions, and ultraprecise measurements of the soliton repetition frequency. 

Our work draws on advances in nanophotonics and photonic-crystal devices that provide access to otherwise challenging or impossible to achieve phenomena. Take, for example, exotic refractive phenomenon\cite{Kocaman2011}, strong light-matter interactions \cite{Miura2014}, and coupling to radiofrequency or phonon modes \cite{Fang2016}. Moreover, photonic structures have been demonstrated to suppress \cite{Petrovich2008} and enhance \cite{Sharma2015} nonlinear effects, engineer small mode volume \cite{Hu2016}, create sophisticated group-velocity dispersion profiles \cite{Kim2017, Moille2018}, realize slow-light effects \cite{McGarveyLechable2017}, and control resonator mode splittings \cite{Lu2014}. Photonic-crystal devices are dielectric structures with sub-wavelength spatial periodicity \cite{Joannopoulos1997} that restrict scattering to discrete momentum values \( k_m=k_0+\frac{2m\pi}{\Lambda}\) not interacting with free-space modes, where $\Lambda$ is the periodicity and $m$ is an integer. In a photonic resonator, the bandgap imposes reflective boundaries to confine light as in a Fabry-Perot cavity \cite{Yu2019}. In our experiments, we use the bandgap instead in an edge-less boundary condition-- a complete ring without edges --to modify a select mode of the PhCR. This condition, combined with an even number of nanopattern periods, frequency-aligns the bandgap to a mode of the PhCR \cite{McGarveyLechable2014}. 

\begin{figure}[htb]
\centering
\fbox{\includegraphics[width=0.95\linewidth]{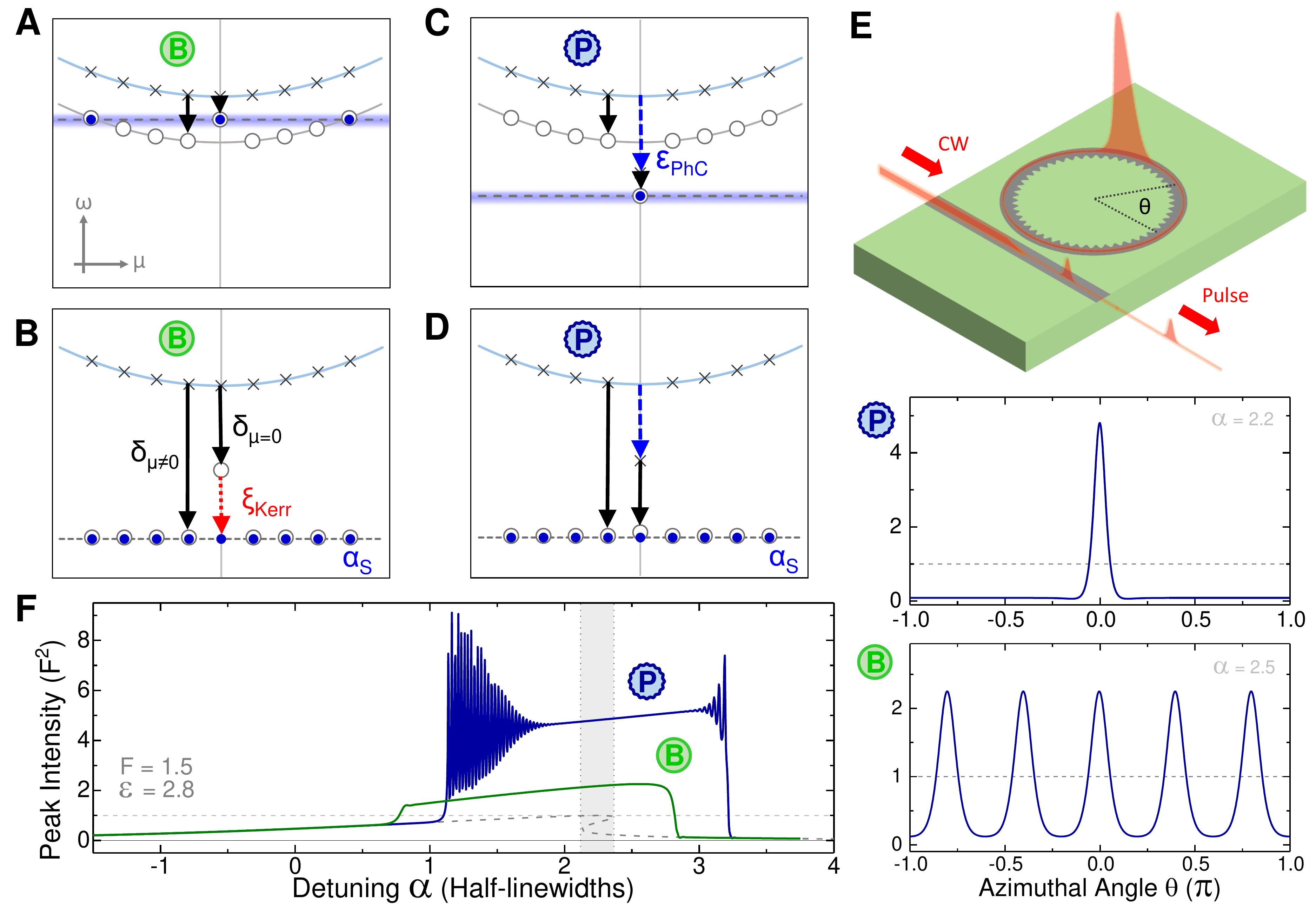}}
\caption{Mode structure for \textcolor{black}{the Kerr} resonator, showing the cold- (cross) and hot-cavity (open circle) resonances, pump laser (dashed line), and light in each mode (solid circle). \textcolor{black}{The left panels show the } (\textbf{A}) Turing pattern, and (\textbf{B}) DKS state in the \protect\includegraphics[width=0.75\baselineskip]{B_logo.png} resonator, with the \textcolor{black}{DKS} Kerr mismatch (dashed red arrow). (\textbf{C}) and (\textbf{D}) show the \protect\includegraphics[width=0.75\baselineskip]{P_logo.png} resonator with photonic crystal shift (dashed blue arrow) at the corresponding Kerr shift, with (\textbf{D}) in the pulse state.  (\textbf{E}) Illustration of optical pulse formation in a photonic ring resonator. (\textbf{F}) Simulated peak power versus pump laser detuning for the \protect\includegraphics[width=0.75\baselineskip]{B_logo.png} (green) and \protect\includegraphics[width=0.75\baselineskip]{P_logo.png} (blue) resonators, with the analytic flat amplitude (dashed gray) for reference. The corresponding intensity profiles are shown in the right panels.}
\label{Fig:Conceptual}
\end{figure}

 
Figure \ref{Fig:Conceptual} introduces the mode-frequency structure of a ring resonator (\BLogo{}) and a PhCR (\PLogo{}), emphasizing how modifying the pump mode affects Turing-pattern and Kerr-soliton generation. The diagrams plot the modal detuning $f_\mu - (f_0 + \mu \cdot FSR )$ for each mode $\mu$, showing the \textit{cold}-resonator modes that correspond to comb modes $\mu$ (crosses) and the \textit{hot}-resonator modes (open circles). The \textit{cold} resonances follow the integrated dispersion \(D_\text{int} = \omega_\mu-\omega_0 - D_1 \mu = 1/2 \, D_2 \, \mu^2 + \epsilon_\text{PhC}\cdot \left( 1-\delta(\mu)\right)\), where $\omega_\mu$ is the angular frequency, $D_1$ is the free-spectral range, $\epsilon_\text{PhC}$ is the frequency shift of the pump mode, and $\delta(\mu)$ is the Kronecker delta function. We additionally shift the \textit{hot} resonances by the Kerr shift $\delta_\mu$, indicating phase accumulation from the Kerr effect. At the onset of flat-state breakup, $\delta_{\mu=0}$ is half that for all other modes. Therefore a natural phase matching exists for FWM to the mode $\mu'$, shown in Fig. \ref{Fig:Conceptual}A where the horizontal dashed line matches the shifted $D_\text{int}$ curve. Hence, the Turing pattern emerges, initially composed of pump and $\pm \mu'$ modes (blue dots). The stationary soliton state (Fig. \ref{Fig:Conceptual}B) of the ring resonator \BLogo{} involves Kerr frequency shifts to balance dispersion across many equidistant comb modes (blue dots); the horizontal line in Fig. \ref{Fig:Conceptual}B indicates the pump laser. However, since the pump-mode Kerr shift is reduced, only large $\alpha$ balances the Kerr mismatch \(\xi_\text{Kerr}\ = \delta_{\mu\neq 0}-\delta_{\mu=0}\) This detuning precludes spontaneous formation of the Turing pattern, but also the formation of solitons, as the low flat state amplitude is below threshold. See \textit{Supplemental} Sect. \ref{S:KerrShift} for more details. 

With a PhCR \PLogo{}, we program the frequency shift $\epsilon_\text{PhC}$ to alleviate the $\xi_\text{Kerr}$ mismatch of the soliton state. The negative shift of both the \textit{cold} and \textit{hot} resonator at comb mode $\mu$ are apparent in Fig. \ref{Fig:Conceptual}C, D. Under this condition, the Turing pattern no longer emerges from the flat state when the pump mode is energized, since the natural FWM phase matching is removed; see the horizontal line in Fig. \ref{Fig:Conceptual}C. Importantly, the shift $\epsilon_\text{PhC}$ moves the \textit{cold} pump mode toward lower frequency by an amount commensurate with the mismatch $\xi_\text{Kerr}$, thereby compensating for the reduced Kerr shift on the pump mode, bringing it approximately onto resonance with the pump laser. Operationally, soliton formation in a PhCR proceeds with the configuration shown in Fig. \ref{Fig:Conceptual}E. We integrate a PhCR device and a coupling waveguide on a silicon chip. The frequency shift $\epsilon_\text{PhC}$ is controlled by the nanopatterning on the ring, while the pump laser field $F$ couples evanescently into the PhCR from a waveguide. The continuous-wave pump laser energizes the PhCR and creates a stable soliton pulse-train at the output.


To verify the physical understanding presented above, we use the LLE to calculate $\psi$ during a sweep of the pump-laser frequency across the pump mode; See \textit{Supplemental} Sect. \ref{S:PSLLE} for an LLE with the mode shift. Figure \ref{Fig:Conceptual}E shows the peak intensity $|\psi|^2$ versus detuning for the ordinary \BLogo{} and photonic-crystal \PLogo{} resonators, respectively. All frequency variables including $\alpha$ and $\epsilon_\text{PhC}$ are in unit of half-width-half-max linewidths unless otherwise specified. Aside from changing $\epsilon_\text{PhC}$ from 0 to 2.8 to activate the PhCR frequency shift, both simulations are performed with the same conditions, namely $F=1.5$, $\beta= -0.17$. The ring \BLogo{} produces the 5-lobe Turing pattern (lower panel) as the pump detuning is swept completely across resonance, corresponding to a range of $\alpha$ from $-2$ to $4$. We then introduce the PhCR case \PLogo{} and carry out the same $\alpha$ sweep. In contrast to the \BLogo{} case, a single pulse forms with abrupt onset. Neither Turing patterns nor chaotic states form during the sweep. Furthermore, the pulse demonstrates two distinct sections of oscillatory stages, known as `breather' soliton states \cite{Kippenberg2018}. The curious reappearance of the breather state at the end of the sweep is also contrasts with \BLogo{} resonator soliton behavior, and we observe this in our experiments.


\begin{figure}[htb]
\centering
\fbox{\includegraphics[width=0.95\linewidth]{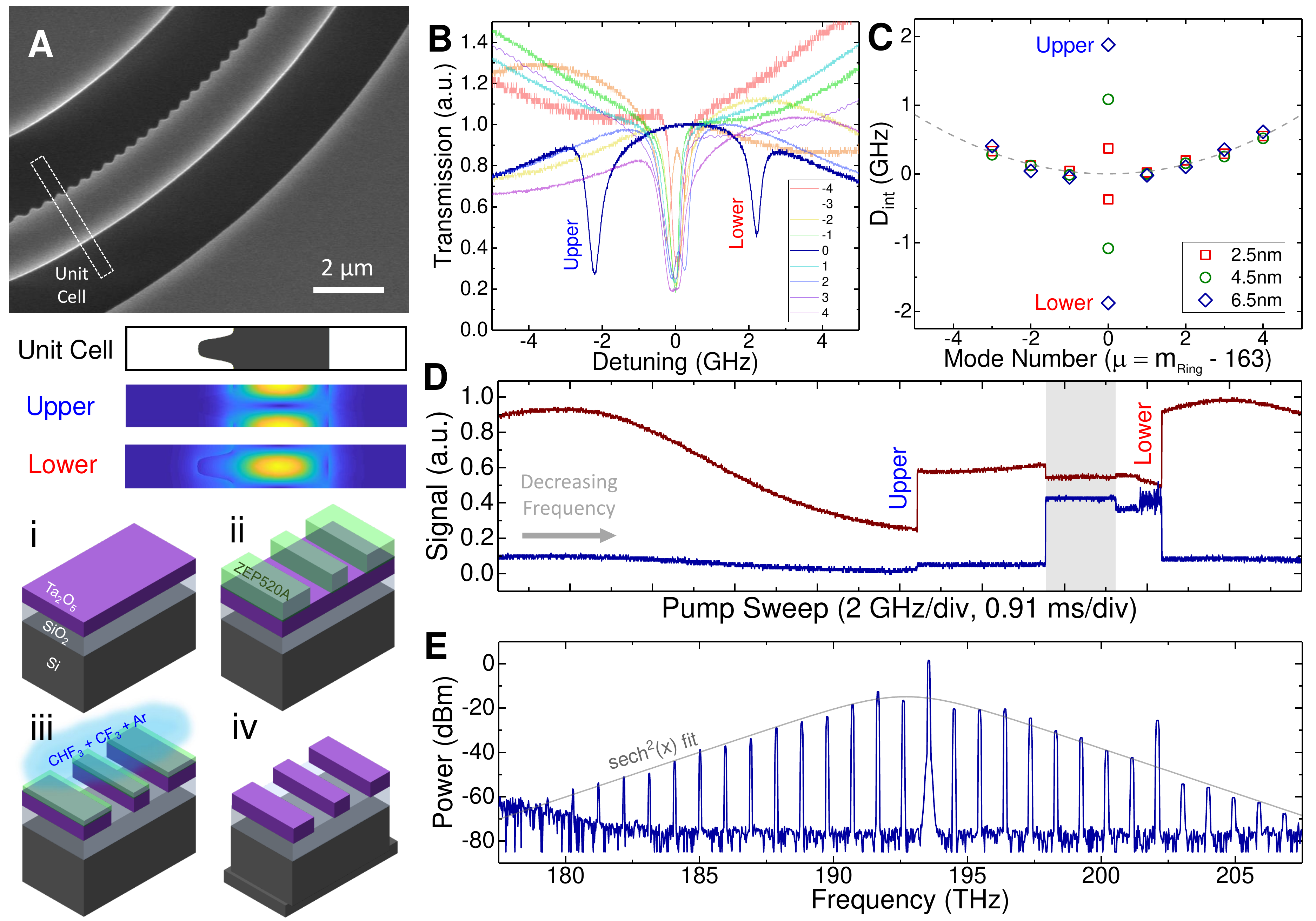}}
\caption{(\textbf{A}) Electron microscope image of the PhCR, showing the unit cell geometry and electric field profiles. Fabrication steps are shown as (\textbf{i}) substrate, (\textbf{ii}) electron-beam lithography, (\textbf{iii}) reactive ion etching, and (\textbf{iv}) dicing. (\textbf{B}) Laser frequency sweep traces of the shifted mode and nearby modes. (\textbf{C}) Mode frequencies for resonators with varying $A_\text{PhC}$, where $\mu=0$ is the $m_\text{Ring}=163$ resonance mode. (\textbf{D}) Laser frequency sweep traces of pump transmission (red) and comb power (blue) demonstrating a spontaneous step, where the stable pulse state is shaded in gray. (\textbf{E}) Optical spectrum of the stable state.}
\label{Fig:DevSteps}
\end{figure}

Figure \ref{Fig:DevSteps} presents our PhCR devices and experimental evidence for spontaneous soliton formation, according to the principles laid out above. We create a PhCR device with the oscillatory nanopattern indicated in Fig. \ref{Fig:DevSteps}A. A unit cell of the pattern is defined by a sinusoidal shape, characterized by the pattern periodicity and peak-to-peak amplitude $A_\text{PhC}$. The periodicity enforces a photonic bandgap that necessarily overlaps one particular PhCR mode, denoted as the pump mode $\mu=0$, in the 1550-nm wavelength range, owing to an equal azimuthal mode number of pattern periods and optical-mode fringes. The bandgap lifts the degeneracy of counter-propagating light in the PhCR, creating modes shifted to higher and lower frequency by an amount $\epsilon_\text{PhC}$. Since the nanopattern is edgeless-- circumferentially uniform -- high resonator $Q$ is maintained. The properties of the other PhCR modes ($\mu \neq 0$) with $\epsilon_\text{PhC} \approx 0$, including nonlinearity and GVD, are preserved under the geometric modification. In particular, the GVD depends sensitively on the thickness and ring-waveguide width (RW) as in an \BLogo{} resonator. We fabricate our devices from a 570-nm-thick tantalum pentoxide (Ta$_2$O$_5$, hereafter tantala) photonics layer \cite{Jung2019}, which is deposited on an oxidized silicon wafer. We use electron-beam lithography to define the photonics pattern for a wafer, and we transfer it to the tantala layer by use of fluorine reactive-ion etching. A final UV lithography process defines several chips on the wafer, and we dry-etch facets in the tantala and oxide layers, and the silicon wafer. See \textit{Supplemental} Sect. \ref{S:Fab} for more details.

In our experiments with PhCRs, we characterize $\epsilon_\text{PhC}$ by spectroscopy measurements. We fabricate up to $\sim75$ PhCRs on a chip with a systematic, few-linewidth variation of $\epsilon_\text{PhC}$ and the waveguide-resonator coupling gap to optimize the conditions for spontaneous soliton formation. To measure $\epsilon_\text{PhC}$, we couple light to and from the chip with a standard lensed-fiber system. Using a 1550-nm tunable laser as input, we record the transmission at the output with a photodetector. Figure \ref{Fig:DevSteps}B presents several PhCR mode resonances in the 1550-nm band, with applied frequency offsets so the resonances coincide, that demonstrate a single mode frequency splitting. We label the non-degenerate modes as upper and lower, with the latter at a setting of $\epsilon_\text{PhC}$ consistent with spontaneous soliton formation. Our experiments focus on gaps for near-critical coupling, and this data indicates a loaded PhCR  $Q$ of $\sim$ 400,000. By adjusting the nanopattern amplitude through our e-beam lithography, we systematically vary $\epsilon_\text{PhC}$; see Fig. \ref{Fig:DevSteps}C. In the range of $A_\text{PhC}$ used in this work, the $Q$ factors are unaffected, compared to \BLogo{} resonators fabricated on the same wafer. With a nanopattern amplitude of only a few nm, we control $\epsilon_\text{PhC}$ for the $\mu=0$ mode, whereas the $\mu'\neq 0$ modes exhibit an anomalous GVD of $D_2= 2\pi\cdot 69.0$ MHz/mode. The results confirm our fabrication process provide the high device geometry resolution and low optical loss to build PhCRs to support the pulses.




We search for spontaneous soliton formation in a PhCR with $\epsilon_\text{PhC}=2.2$ by sweeping the frequency of the pump laser with $\sim 36$ mW of on-chip power; Fig. \ref{Fig:DevSteps}D presents a $\sim20$ GHz sweep range from high to low frequency that spans the upper and lower resonances. With photodetectors, we monitor both transmission through the PhCR device (red trace)  and the power of generated comb modes (blue trace), which we obtain by filtering out the pump. These data show the presence of thermal bistability effects, which distort the resonances into a triangle shape, and the effects of nonlinear comb generation. In particular, we observe no comb power at the upper resonance, as the upper mode is shifted away from the $\mu'$ modes needed for FWM. Whereas at the lower resonance we observe immediate comb formation, corresponding to the step change in comb power that agrees with our simulation in Fig. \ref{Fig:Conceptual}F. We assess that this nonlinear state on the lower resonance, indicated by the shaded range in Fig. \ref{Fig:DevSteps}D, is a dissipative Kerr soliton that spontaneously forms under certain conditions of pump power and laser detuning. Additionally, we observe a nonlinear state on the lower resonance that exhibits relatively higher comb power variance, likely a breather state as indicated theoretically in Fig. \ref{Fig:Conceptual}F. The breather state at higher detuning than the stable state suggests a modified optical state phase diagram yet to be explored. Operationally, we adjust the pump power to maximize the pump-frequency existence range of the low-noise spontaneous soliton step, and we hand adjust the laser frequency into this range. Under these conditions we record the optical spectrum (Fig. \ref{Fig:DevSteps}E) of the soliton comb, which exhibits a clear $sech^2(\nu)$ profile as shown by the gray line. The remainder of our paper presents measurements of such spontaneous solitons.

We attribute the ease of spontaneous-soliton capture to desirable thermal behaviors of the PhCR. Conventionally in a \BLogo{} device, capturing and sustaining a soliton is difficult as a result of rapid heating and cooling of the microresonator \cite{Stone2018, Brasch2016kick}. Soliton initiation in the \BLogo{} resonator under CW excitation is proceeded by Turing patterns or chaotic states, which are multiple-pulse states with high average intensity. Conversely, the desired soliton state is a single pulse with a relatively low average intensity. Hence, the root of thermal instability is the transition of nonlinear state in a microresonator. The \PLogo{}-resonator spontaneous solitons offer two primary advantages: First, in soliton initiation, we bypass the high average intensity states and avoid their heating effects to the resonator. Second, we keep the pump laser on-resonance in the soliton state (note the drop in transmission trace in Fig. \ref{Fig:DevSteps}D as the pulse forms, indicating a more resonant condition), therefore minimizing changes to the in-resonator pump amplitude as the soliton forms. Together, these factors minimize the intensity changes in the PhCR, allowing pulses capturing by hand-tuning alone.



\begin{figure}[htb]
\centering
\fbox{\includegraphics[width=0.95\linewidth]{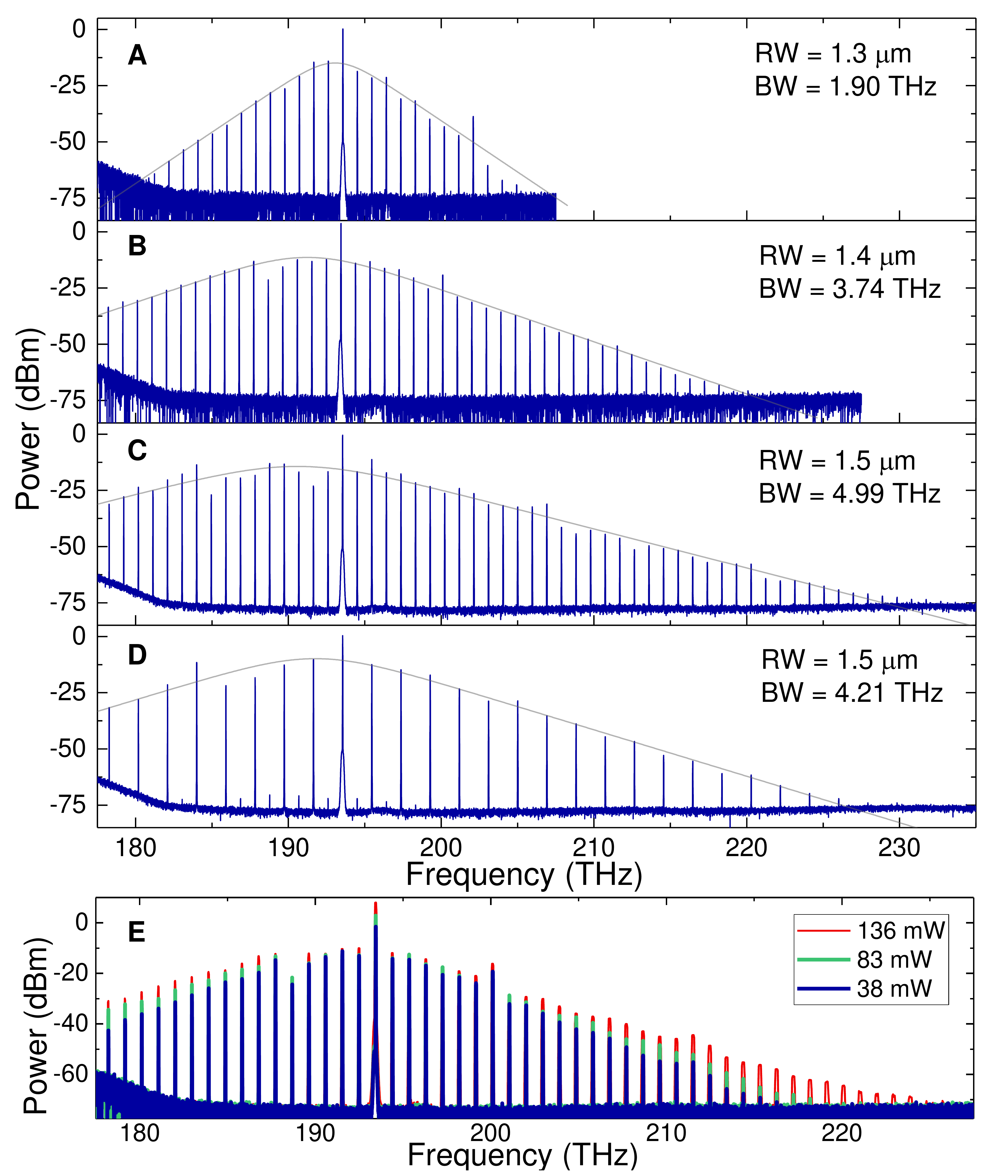}}
\caption{Optical spectra from \textcolor{black}{PhCR}s with average ring width (\textbf{A-C}) 1.3, 1.4, and 1.5 $\mu$m. (\textbf{D}) The two-pulse state on the 1.5 $\mu$m device. The gray traces are fits with the form $y = A_0+ 10 log(sech^2((x-x_0)/BW))$.  (\textbf{E}) Spectra versus power for the RW = 1.4 $\mu$m device.}
\label{Fig:Spectra}
\end{figure}

To explore the universality of spontaneous-soliton formation, we demonstrate soliton bandwidth control by tuning the GVD of the PhCR and the pump-laser power. We control the GVD directly by varying the RW from 1.3 to 1.5 $\mu$m, providing decreasing anomalous GVD that we can understand from FEM calculation of the PhCR resonator mode structure. Based on the LLE, this change should affect an increasing soliton bandwidth. We tune by hand into the soliton states on these devices and acquire their optical spectra, plotted in Fig. \ref{Fig:Spectra}A-C. The spectrum bandwidth broadens with decreasing anomalous GVD as expected. Interestingly, we acquired a stable two-pulse state at lower detuning on the RW = 1.5 $\mu$m device, shown in Fig. \ref{Fig:Spectra}D. The two-pulse state suggests that the parameter space of the PhCR -- an interplay between dispersion and mode shift -- supports more steady states beyond the single spontaneous pulse. We also varied the pump laser power for the RW = 1.4 $\mu$m device, see Fig. \ref{Fig:Spectra}E, resulting in widening of the spectral envelope consistent with the DKS. However, unlike the conventional case where increasing pump power monotonically lengthens the soliton existence range \cite{Guo2017}, the PhCR produces strong breather states at high power. More study is underway to fully explore this behavior.


\begin{figure}[htb]
\centering
\fbox{\includegraphics[width=0.95\linewidth]{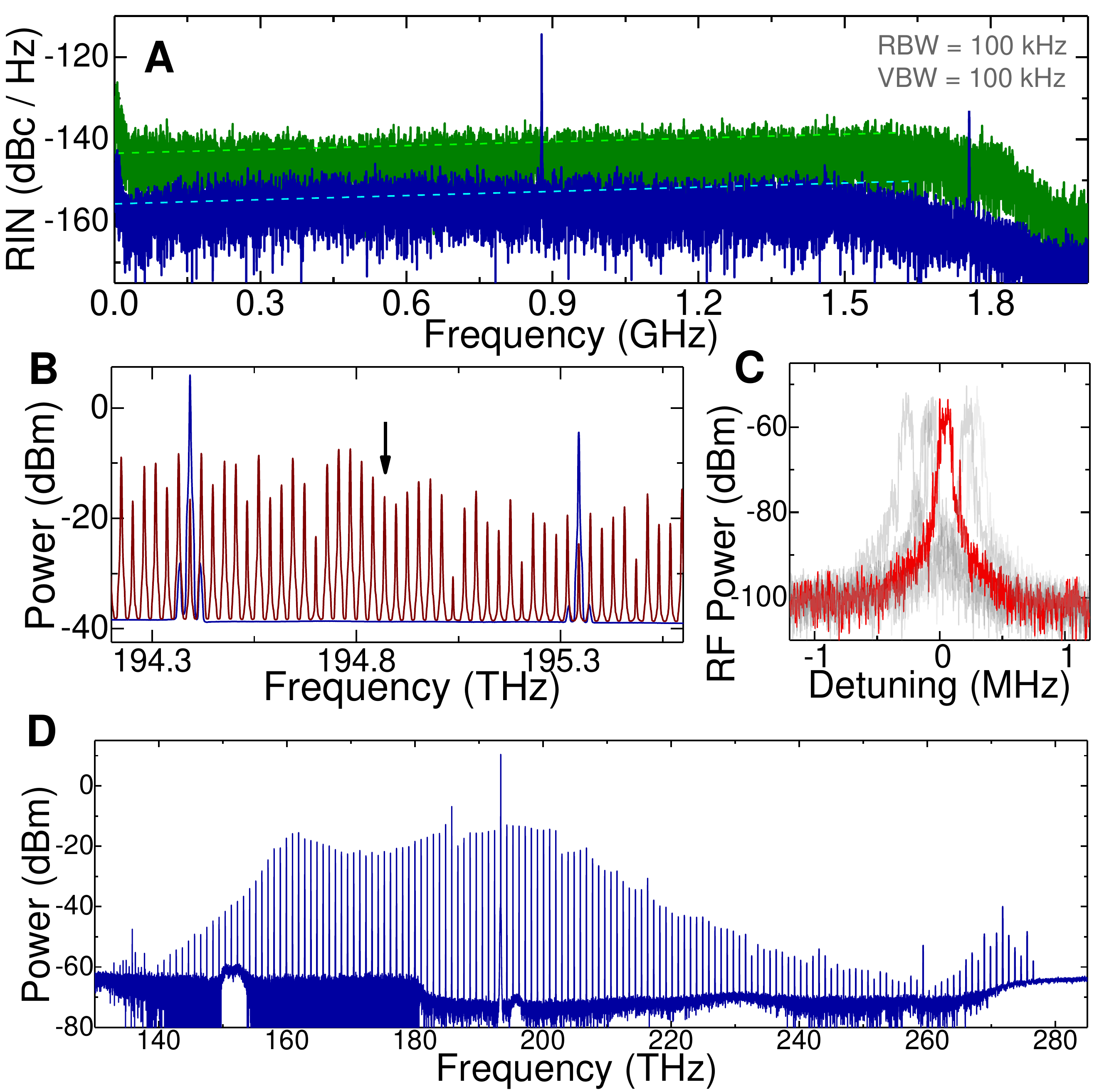}}
\caption{(\textbf{A}) The relative intensity noise on the comb power of a breather (blue) and quiet (green) states. The dash lines show the detector noise floor corresponding to the carrier power. (\textbf{B}) Bridging between the 1 THz separation between comb lines (blue) by creating sidebands (red) using electro-optic modulation. The arrow indicates the overlapping mode. (\textbf{C}) Electronic beatnote from the overlapping mode. The gray traces show five consecutive measurements. (\textbf{D}) Optical spectrum of broad-band soliton.}
\label{Fig:Coherence}
\end{figure}
Stationary microresonator solitons output an optical pulse-train with fixed period, which composes a low-noise, equidistant frequency comb suitable for optical-frequency measurements \cite{Briles2018, Drake2019}. Therefore, verifying the spectral-noise properties of spontaneous solitons in PhCR is of utmost importance. Figure \ref{Fig:Coherence} presents intensity- and frequency-noise measurements, excluding the pump laser, of a spontaneous soliton, which we generate in a device with RW = 1.4 $\mu$m, $\epsilon_\text{PhC}$=3.0. The relative intensity noise (RIN, Fig. \ref{Fig:Coherence}A) of a stationary soliton and a breather soliton is below $-140$ dBc/Hz over a Fourier frequency range to 1.8 GHz. Here, the photodetected soliton power is 282 $\mu$W and the spur-free dynamic range is excellent, whereas the breather state manifests a single peak at 878 MHz and supports higher power and hence lower RIN. These measurements are currently limited by the comb power and the detector noise floor.

To measure the $\sim$ 1 THz PhCR soliton repetition frequency, we apply electro-optic (EO) phase modulation to create a low-frequency heterodyne beat between two soliton comb modes \cite{Drake2019}; the optical spectrum trace in Fig. \ref{Fig:Coherence}B indicates the soliton modes in blue and the EO  sidebands in red. We choose the EO drive frequency such that the $\pm17^ \text{th}$ order sidebands (arrow in Fig. \ref{Fig:Coherence}B) generate an optical heterodyne on a photodetector, after filtering out that pair. We identify the tone thus generated as the heterodyne, as it varies with the EO drive frequency at $34.2$ MHz/MHz  in agreement with the sideband orders. We present the heterodyne spectrum in Fig. \ref{Fig:Coherence}C, which shows the typical lineshape with $\sim 50$ kHz linewidth and $<1$ MHz fluctuations. We attribute these properties to thermal noise \cite{Drake2019cooling} and thermal drift of the microresonator. Finally, we demonstrate a PhCR device with optimized dispersion to create a spontaneous DKS with near-octave bandwidth, shown in Fig. \ref{Fig:Coherence}D. The $F^2$ value for this trace is estimated to be 8.7, normalized to threshold power of $\mu=\pm 1$ modes. We anticipate these optimized devices to enable f-2f self referencing in the future.

In conclusion, we have presented spontaneous and deterministic generation of Kerr solitons in edge-less PhCRs, enabled by compensating the Kerr shift mismatch between the pulse state and its pump mode. Mode-shifting by nanopatterning enables spontaneous generation, whereas we retain the capability to engineer broadband dispersion with the bulk ring geometry. The importance of the nanophotonic capabilities presented in this work is two-fold: First, the ability to controllably shift modes while maintaining the bulk dispersion profile provides a tool to explore the physics occurring in a nonlinear process. Here, the capability modifies the behavior of the pump mode, but we envision applications such as direct engineering of dispersive waves \cite{Matsko2016} or soliton crystals \cite{Cole2017}, potentially enabling inverse design methods for arbitrary desired waveforms; Second, the spontaneous formation nature of pulses demonstrated here significantly reduces the system complexity for a soliton formation and stabilization system, enabling low power consumption, packaging-friendly devices, or integrated systems with multiple independent pulse sources. We envision spontaneous pulse devices like the PhCRs presented in this work to become building blocks for future nonlinear optics and integrated-photonics technologies.

\bibliography{scibib}

\section*{Acknowledgments}
Funding provided by the DARPA DODOS, DRINQS, and PIPES programs. We acknowledge the Boulder Microfabrication Facility, where the devices were fabricated. We thank Travis Briles and Jeff Chiles for a careful reading of the manuscript. This work is a contribution of the U.S. Government and is not subject to copyright. Mention of specific companies or trade names is for scientific communication only, and does not constitute an endorsement by NIST.

\section*{Supplementary materials}
Materials and Methods\\
Design and Fabrication\\
Derivation of Modified LLE\\
Kerr Shift Calculation\\
Pulse Formation Dynamics\\
Figs. S1 to S2\\
References \textit{(37-38)}

\clearpage

\setcounter{figure}{0} 

\onecolumngrid
{\centering

\Large
\textbf{Supplenmental Materials for: Spontaneous Pulse Formation in Edge-Less Photonic Crystal Resonators}

}
\vspace{15pt}
{\centering

Su-Peng Yu$^{1,2\ast}$, Daniel C. Cole$^{1,2}$, Hojoong Jung$^{1,2}$, Gregory T. Moille$^{3,4}$,

Kartik Srinivasan$^{3,4}$, and Scott B. Papp$^{1,2}$

}
\vspace{10pt}
{\centering

\normalsize{$^{1}$Time and Frequency Division, NIST, Boulder, CO 80305, USA}

\normalsize{$^{2}$ Department of Physics, University of Colorado, Boulder, CO, 80309, USA}

\normalsize{$^{3}$ Microsystems and Nanotechnology Division, NIST, Gaithersburg, MD 20899, USA}

\normalsize{$^{4}$ Joint Quantum Institute, NIST / University of Maryland, College Park, MD 20742, USA}

}

\vspace{10pt}

\twocolumngrid

\section{Materials and Methods} \label{S:MatMtd} 
\begin{figure}[htb]
\centering
\fbox{\includegraphics[width=1.0\linewidth]{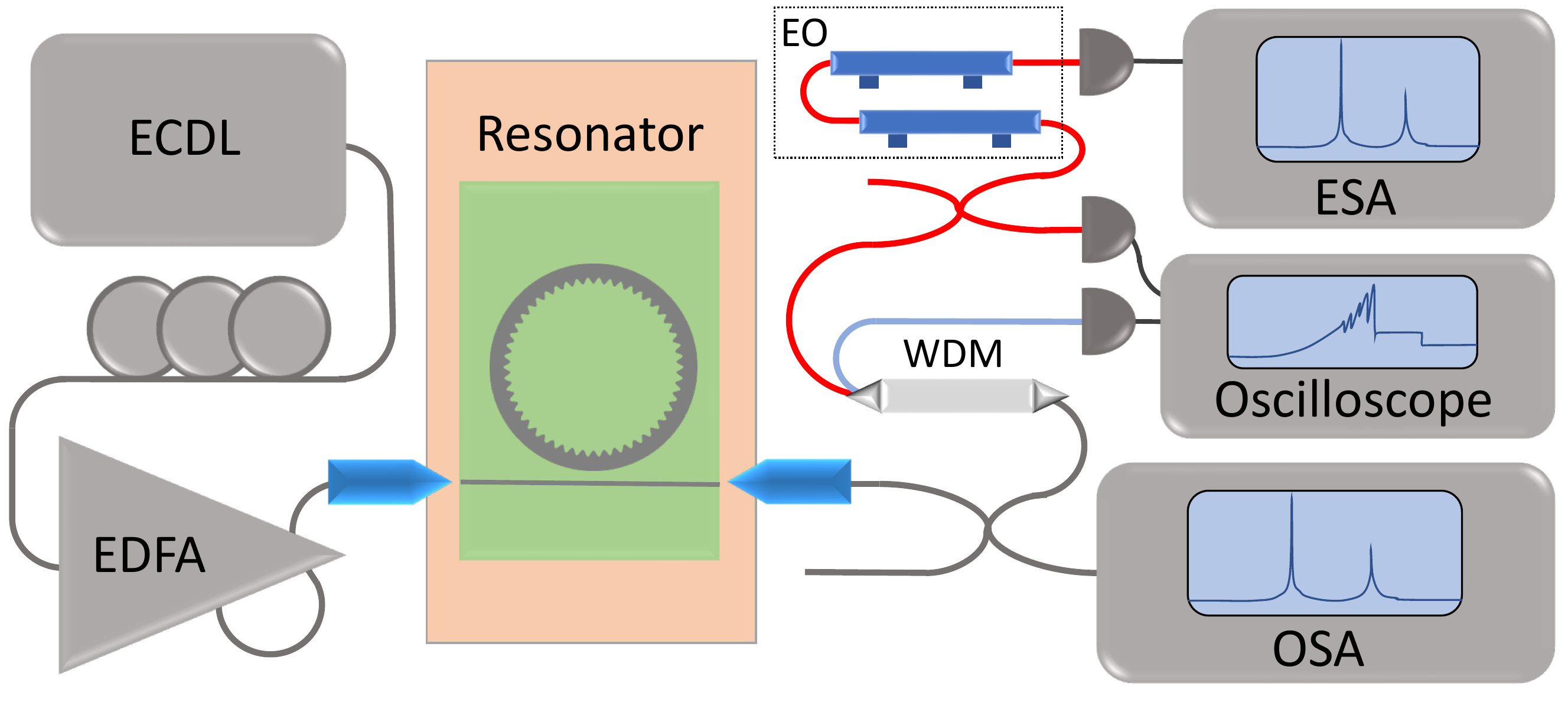}}
\caption{Illustration of the optical testing setup.}
\label{Fig:System}
\end{figure}

Here we provide details for the optical setup used in this work, illustrated in Figure S\ref{Fig:System}. The light source is a C-band tunable external-cavity diode laser (ECDL) with fiber-coupled output. The light goes through a fiber isolator and then to a set of fiber polarization controllers. A 90\% fused fiber coupler is added between the laser and the polarization controller to tap the laser light for a Mach-Zehnder interferometer and a \textcolor{black}{wavelength meter (wavemeter, 40 MHz resolution)} for frequency measurements. We use the wavemeter to precisely measure modes frequencies within the ECDL tuning range, enabling us to characterize the dispersion of PhCRs. For comb generation experiments, the laser is amplified using an erbium-doped fiber amplifier (EDFA), with a tunable band-pass filter to suppress the amplified spontaneous emission of the EDFA for cleaner spectra. For passive measurements, the EDFA and the filter are bypassed. We send the light into the photonic chip using a lens fiber mounted on a three-axis flexure stage, controlled by manual micrometers. The damage threshold of our devices is typically above 1 W incident power. The typical coupling efficiency between fiber and chip is $\sim$25\% per facet, limited by the mode mismatch between the air-clad waveguides and lens fibers. The chip is placed on a copper block for thermal contact. The output is collected with another lens fiber on translation stage. For passive measurements, we measure the outcoming power using an amplified photodetector, plotting the transmission versus frequency on an oscilloscope.

During the comb generation experiments, we continuously monitor a portion of the outcoupled light with an OSA. With photodetectors that have 150 MHz bandwidth, we also monitor the pump-laser transmission of the resonator and the comb power, which we obtain by filter out the pump contribution. The comb power signal provides critical information on break up of the flat background and soliton initiation, and for monitoring the intensity-noise level of soliton states.
To diagnose breather soliton oscillations and perform intensity-noise measurements, we use a high-speed photodetector \textcolor{black}{(1.6 GHz bandwid)} and a electronic spectrum analyzer. 

The comb-power channel, after filtering out the pump, is also used for the beatnote measurements. We pass the comb light through two cascaded EO phase modulators, driven far above $V_\pi$ to introduce multiple sidebands to span the 1 THz frequency spacing between the comb lines, shown in main text Figure \ref{Fig:Coherence}B. We choose the EO modulation frequency to be 28.000 GHz so the $\pm$17th sidebands from adjacent comb lines will come into close vicinity. To improve the signal to noise ratio for the beatnote measurements, we amplify the EO output with a semiconductor optical amplifier and select the overlapping modes using a \textcolor{black}{tunable optical filter with } a 50GHz passband.

\section{Design and Fabrication} \label{S:Fab}

Here we describe the design process to create a PhCR. We calculate the \includegraphics[width=0.75\baselineskip]{B_logo.png} resonator dispersion and photonic bandgap using a finite-element method program. The dispersion calculation yields the propagation constant $k_\text{eff}$ for each RW, ring radius R, and frequency. The azimuthal mode order m of the PhC is then calculated by the boundary condition $k_\text{eff} \cdot 2\pi R = 2m\pi$. The PhCR modulation is then introduced with the periodicity $2\pi R/2m$ and sinusoidal peak-to-peak amplitude $A_{PhC}$ on the interior of the ring. The sinusoidal shape is chosen as it can be fabricated reliably to very small amplitude using lithography and plasma etching. A bus waveguide approaches the smooth outer edge of the resonators. The strength of the evanescent coupling between the resonator and the bus is controlled by the gap between the two. On the edges of the chips where the bus waveguides terminate, the waveguides are inversely tapered to improve mode-matching to lens fibers. We generated the mask files using a \textcolor{black}{pattern-defining} script and the \textit{CNST Nanolithography Toolbox} \cite{CNSTBox}. Typically, we place up to 70 PhCRs and their bus waveguides per chip in an evenly spaced array. Fine sweeps of $A_{PhC}$ and coupling gap are included to achieve the correct mode shifts and near-critical coupling.

The fabrication procedure of our devices is as follows: We obtain 3-inch silicon wafers with 380 $\mu$m thickness and 3 $\mu$m thermal silicon dioxide on both sides. The tantala device layer is deposited onto the wafer to 570 nm thickness by \textcolor{black}{an external supplier}. For lithography, we carry out a double spin-coating of \textit{ZEP520A} resist to reach a total resist thickness of 1 $\mu$m, then expose the resist in electron beam lithography (EBL) \textcolor{black}{operating} at 100 kV. All device patterns are defined on this EBL step. We develop the resist and transfer the pattern using plasma etching with an \textcolor{black}{inductively coupled plasma etching} tool, and a $CHF_3 + CF_4 + Ar$ chemistry. The ratio between $CHF_3$ and $CF_4$ is varied to achieve vertical sidewall, while the Ar gas was found to improve sidewall smoothness. The etch selectivity is sufficient to clear the device layer with the resist thickness used. A dicing pattern is put onto the wafer using UV lithography and the \textit{SPR-220} photoresist. We etch through the bottom thermal oxide layer using a plasma etch with $CHF_3 + O_2$ chemistry. The resist is stripped using solvents, and the UV lithography step is carried out again for the deep-RIE dicing using the $C_4F_8+SF_6$ chemistry. We then clean the wafer of the fluoro-polymer deposited during the RIE steps using \textit{DuPont EKC265} solvent, followed by a \textit{Cyantek Nanostrip} soak for final cleaning. The chips are then mechanically removed from the wafer and are ready for testing.

\section{Derivation of Modified LLE} \label{S:PSLLE}

Here we provide a modified LLE to accommodate the influences of the shifted pump mode. Importantly, the form of the modified LLE admits the steady-state solutions of the LLE, with a effective pump field reflecting the influence of the shifted mode. We begin with the LLE in the modal basis,
\begin{equation}
\partial_\tau a_\mu = -(1+i\alpha)a_\mu + \frac{i}{2}\beta \mu^2 a_\mu + \delta_{\mu 0} F+ \hat{\mathcal{F}}\{i|\psi(\theta)|^2\psi(\theta)\}
\end{equation}
where $a_\mu$ is the field amplitude in mode $\mu$, $\beta=-\frac{2}{\kappa}D_2$ stands for the second-order dispersion normalized to linewidth $\kappa$, $\delta_{\mu 0}$ the Kronecker delta function $\delta_{0 0}=1$, zero otherwise, and $\hat{\mathcal{F}}$ the Fourier transform. We generalize the equation to arbitrary dispersion profiles by identifying $\frac{i}{2}\beta \mu^2=-\frac{2i}{\kappa}\cdot\frac{1}{2}D_2 \mu^2$ to be $-\frac{2i}{\kappa}D_\text{int}(\mu)$, where $D_\text{int}(\mu)$ is the integrated dispersion. We implement the pump mode shift with an additional term to the total dispersion:
\begin{equation}
D_\text{int}^{Shifted}(\mu)=D_\text{int}^{Base}(\mu)+\Xi (1-\delta_{\mu 0})
\end{equation}
\noindent where a constant shift of strength $\Xi$ is applied to all modes except zero, so that the zero of detuning $\alpha$ remains defined on the pump mode. A red-shift of the pump mode is associated with $\Xi>0$. The equation becomes:
\vspace{-10pt}
\begin{multline}
\partial_\tau a_\mu = -(1+i\alpha)a_\mu + \delta_{\mu 0} F+ \hat{\mathcal{F}}\{i|\psi(\theta)|^2\psi(\theta)\}\\ -\frac{2i}{\kappa}\left(D_\text{int}^{Base}(\mu)+\Xi (1-\delta_{\mu 0})\right) a_\mu
\end{multline}

\noindent now carry out inverse Fourier transform, under the normalization that $\hat{\mathcal{F}}^{-1}(\delta_{\mu 0})=1$, and that $\hat{\mathcal{F}}^{-1}(\delta_{\mu 0}\cdot a_\mu)=\frac{1}{2\pi}\oint\psi(\theta)d\theta := \bar{\psi}$, we get the pump-shifted LLE in the time domain:
\vspace{-10pt}
\begin{multline}
\partial_\tau \psi(\theta) = -\left(1+i(\alpha+\epsilon)\right)\psi(\theta) - \frac{i}{2}\beta \partial_\theta^2 \psi(\theta) \\+ i|\psi(\theta)|^2\psi(\theta) +F +i\epsilon \bar{\psi}
\end{multline}

where $\epsilon = \frac{2\Xi}{\kappa}$ is the normalized mode shift. We make two observations from the shifted LLE formula: First, in the case of an amplitude that is a constant in $\theta$, $\bar{\psi}=\psi$ and the shift terms cancel, indicating that the resonator responses identically to the unmodified LLE prior to pattern generation. Second, assuming a time-stationary pattern $\psi$ is formed, the $\bar{\psi}$ term is constant in the resonator, and the equation can be interpreted as an LLE with modified parameters:
$$\alpha'=\alpha+\epsilon\qquad F'=|F+i\epsilon \bar{\psi}|$$
This is to say any stationary-state solutions $\psi$ of the modified LLE with parameters $F,\alpha$ also satisfies the LLE with parameters $F',\alpha'$, the later include Turing patterns and Kerr solitons. This equivalence enables the pump-shifted LLE to produce Kerr solitons.

\section{Kerr Shift Calculation} \label{S:KerrShift}



We present an interpretation of the Kerr shift term in the modal basis. Under this interpretation, each of the resonator mode $\mu$ behaves as a nonlinear harmonic resonator, therefore giving physical meanings to the \textit{hot}-resonator modes in the main text. We begin by recalling the case of a single-mode nonlinear oscillator:

\begin{equation}
\partial_t a = -i \omega' a - \gamma_0  a + F e^{i \omega t}
\end{equation}

\noindent where the resonance frequency $\omega'=\omega_0-g |a|^2$ depends on the field amplitude $|a|^2$ through nonlinear coefficient $g$. We identify this cubic term as the Kerr term $i g |a|^2 a$ which results from the resonance frequency change induced by the field amplitude. In the case of the LLE, we start with the Kerr term instead, and assign an inferred modal frequency for each field component $a_\mu$ by casting the time-evolution of the mode in the form of a harmonic oscillator:

\begin{equation}
\partial_\tau a_\mu = -i\alpha  a_\mu + i\delta_\mu \cdot a_\mu + g_\mu\cdot a_\mu + \delta_{\mu 0} F
\end{equation}

\noindent where the detuning $\alpha$ was chosen so the pump field $F$ is not time-dependent in the rotating frame, and $\delta_\mu$ and $g_\mu$ can depend on the in-resonator field profile. This form enables us to identify the modal frequency and gain from the instantaneous rate of change of the phase and amplitude induced by the Kerr effect. We calculate these rates by Fourier transforming the Kerr term \(|\psi|^2 \psi\) for a given field profile $\psi(\theta)$:

$$ \partial_\tau a_\mu\vert_{Kerr} = \hat{ \mathcal{F}} \{i|\psi(\theta)|^2\psi(\theta)\},_\mu$$

\noindent where the subscript $\mu$ for the Fourier transform specifies the $\mu$-th component. Casting this into the harmonic oscillator form, we get:

\begin{equation}
\delta_\mu^{Kerr}=\mathcal{R}\textit{e}\left(\hat{ \mathcal{F}} \{|\psi(\theta)|^2\psi(\theta)\},_\mu /a_\mu \right)
\end{equation}

\begin{equation}
g_\mu^{Kerr}=-\mathcal{I}\textit{m}\left(\hat{ \mathcal{F}} \{|\psi(\theta)|^2\psi(\theta)\},_\mu /a_\mu \right)
\end{equation}

\noindent where $\delta_\mu^{Kerr}$ and $g_\mu^{Kerr}$ are the modal Kerr shift and induced gain on mode $\mu$.

An example for this formula is the modal frequency behavior near the Turing pattern onset threshold. In order to extract the frequency of the modes, we assume the modal fields in the resonator take the form:
$$ \psi(\theta) = a_0 + \eta \cdot u_{\mu'}(\theta) $$
\noindent where $a_0$ is the pump mode amplitude, a constant in the resonator, and $a_{\mu'}=\eta$ an infinitesimal field amplitude in the $\mu'$-th mode, $u_{\mu'}(\theta) = exp(i \mu' \theta)$ is the basis function in the $\mu'$-th mode. To obtain the pump mode shift, we evaluate the Kerr shift term $|\psi|^2\psi$ to zeroth order in $\eta$. Since $a_0$ is a constant over $\theta$, the form trivially gives:

\begin{equation}
\delta_0^{Kerr}=\mathcal{R}\textit{e}(\mathcal{F}\{ |a_0|^2 a_0\},_0/a_0) = |a_0|^2
\end{equation}

\noindent which is just the pump mode intensity. To get the shift for the $\mu'$-th mode, we evaluate the Kerr shift term to first order in $\eta$:
$$|\psi|^2\psi = (|a_0|^2 + \eta\cdot(a_0 u_{\mu'}^*+a_0^* u_{\mu'})+ \mathcal{O}(\eta^2))\cdot(a_0+\eta u_{\mu'})$$
$$= |a_0|^2a_0+2|a_0|^2\cdot \eta u_{\mu'} + a_0^2\cdot \eta u_{\mu'}^* + \mathcal{O}(\eta^2)$$
\noindent where $u_{\mu'}^*=u_{-\mu'}$. Fourier transforming this expression and taking the $\mu'$-th component, only the term with $u_{\mu'}$ is non-vanishing. We get:

\begin{equation}
\delta_{\mu'}^{Kerr}=\mathcal{R}\textit{e}(2|a_0|^2\cdot\eta/\eta)=2|a_0|^2
\end{equation}

\noindent which is twice the shift compared to the pump mode, in agreement to the form in \cite{FactorOfTwoBook}.

\begin{figure*}[htb]
\centering
\fbox{\includegraphics[width=0.95\linewidth]{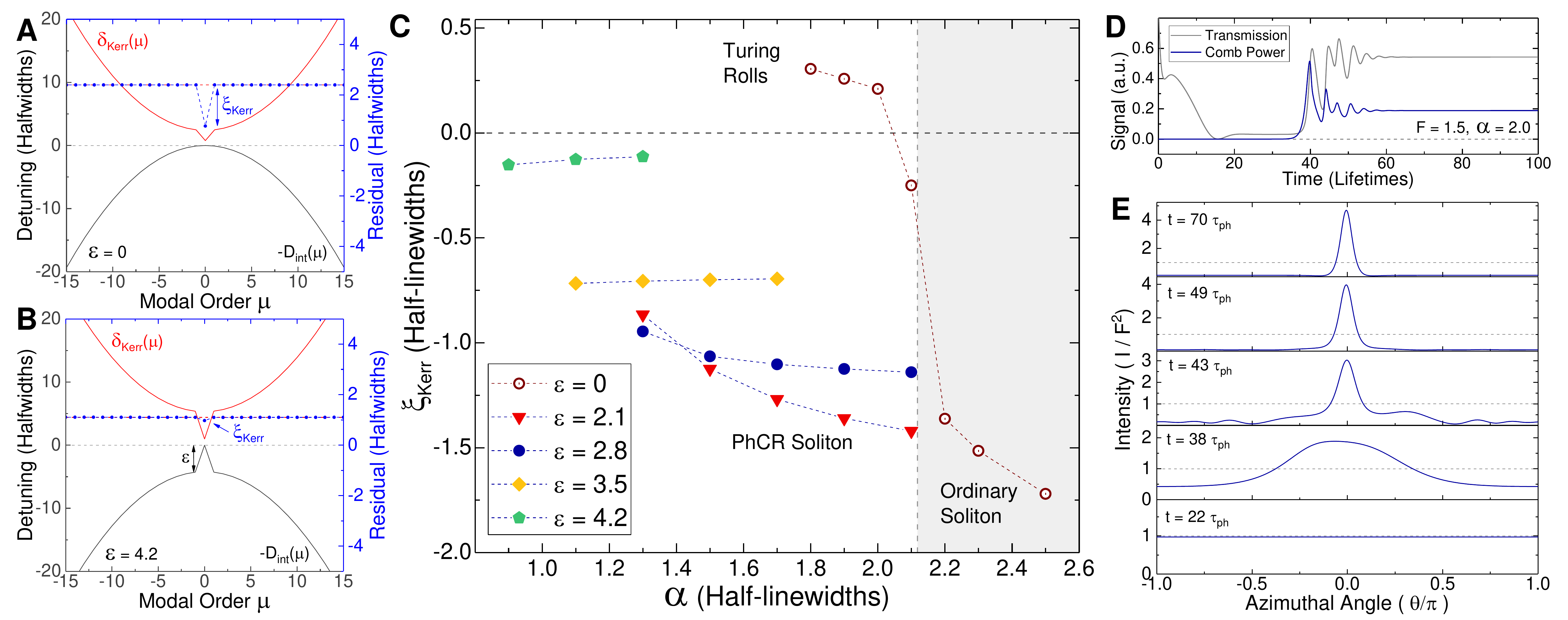}}
\caption{
The balancing of Kerr shift (red) and dispersion (black) for (\textbf{A}) the \protect\includegraphics[width=0.75\baselineskip]{B_logo.png} and (\textbf{B}) \protect\includegraphics[width=0.75\baselineskip]{P_logo.png} soliton pulses, where the sum of the two is magnified and shown in blue. (\textbf{C}) Calculated Kerr mismatch for various mode shift values and pump detuning. (\textbf{D}) Simulated time traces and (\textbf{E}) intensity plots during the spontaneous pulse generation process.}
\label{Fig:Stability}
\end{figure*}

We are now equipped with the theoretical tools to study the Kerr balancing for the soliton states. We generate the soliton field profile using the LLE for sets of parameters $(F, \alpha, \epsilon)$, then calculate the dispersion balancing conditions for each mode, namely $-D_\text{int}(\mu)+\delta_\mu$. We find the term equals $\alpha$ for all non-pump modes, while the gain terms equal 1 to balance loss. The balancing effect enables a stationary, time-independent waveform in the reference frame of the LLE. In the case $\epsilon=0$, the balance is not achieved for the pump mode, but the mismatch is compensated by the pump field $F$ in a manner similar to the forced harmonic oscillator. To study the Kerr mismatch in response to the shifted pump mode, we carry out LLE simulations to obtain the field profiles of the stable pulse states in a resonator for some $\epsilon >0$, and calculate the Kerr shifts for each mode. Kerr shift and $D_\text{int}$ plots for the cases with $\epsilon$=0 and $\epsilon$=4.2 are shown in Figure S\ref{Fig:Stability}A and B. The blue dots show the sum of the two, balancing to the horizontal lines at the value of $\alpha$, except for the pump mode, in agreement with Ref. \cite{Bao2014}. A pronounced Kerr mismatch $\xi_{Kerr}$ is observed for the $\epsilon$=0 case, but is suppressed in the $\epsilon$=4.2 case. We carry out the calculations for intermediate values of $\epsilon$, shown in Fig. S\ref{Fig:Stability}C. The increasing of $\epsilon$ resulted in a gradual reduction of mismatch $\xi_{Kerr}$, down to approximately one quarter of a linewidth. Figure S\ref{Fig:Stability}C also shows the $\alpha$ ranges where the soliton state is stable for each $\epsilon$. We observe that the soliton is stable in detuning ranges in the single-stability range for the given $F$ value for the shifted-pump cases, while in the $\epsilon = 0$ case the soliton is only stable in the bistability range (shaded area in Fig. S\ref{Fig:Stability}C). This leads to the difference that the flat state is stable on the lower branch of the bistability in the $\epsilon = 0$ case, versus the spontaneous generation of patterns from the flat state in the shifted-pump case. Initiating the $\epsilon = 0$ case with a pulse in the single-stability range results in the mode reverting spontaneously to multiple-pulse Turing patterns. This suggests that the mode shift modifies the phase diagram \cite{Godey2014} to enable stable soliton states in ranges where the flat background amplitude is unstable.

\section{Pulse Formation Dynamics} \label{S:PulseForm}

We show the time-evolution of the spontaneous generation of a single pulse in Fig. S\ref{Fig:Stability}D-E. Here the pulse arises spontaneously from the flat state with constant pump $F$ and detuning $\alpha$, seeded only by the vacuum fluctuation. The LLE simulation of pulse generation shows several transient states the resonator goes through to arrive at the DKS state. In this simulation, the resonator is initiated with zero amplitude, and is energized with a fixed pump field F at fixed detuning $\alpha$ for some time until the pulse state stabilizes. We identify four transient states in the pulse generation, shown in Fig. S\ref{Fig:Stability}E, starting from its bottom panel: First, the flat amplitude energizes without producing comb power, until it is sufficiently large that the flat state becomes unstable. Unlike the conventional resonator where the FWM condition can be reached by the large mode density near the pump mode to form Turing patterns order $\mu'$ determined by dispersion, the phase matching is prohibited by the shifted pump mode. Second, with the shifted pump mode, the PhCR instead make a one-lobe sinusoidal pattern once the flat amplitude is sufficiently high. This can be intuitively understood by drawing a quadratic curve across the three modes $\mu=0, \mu'=\pm 1$ for the PhCR mode structure. The high positive curvature of this curve affects a local strong anomalous dispersion, causing a transient Turing pattern of order $\mu'=1$ to form. This transient pattern breaks the $\theta$-symmetry in the resonator, seeding the resonator for a single pulse. Third, the one-lobe pattern begins to sharpen. This is because unlike a true high-anomalous-dispersion resonator where $\mu'>\pm1$ modes are FWM-mismatched from strong dispersion, the $\mu'>\pm1$ modes of the PhCR follow the base dispersion, therefore are sufficiently phase-matched and can be energized. The energizing of $\mu'>\pm1$ modes lead to sharpening of the peak in time domain, and the broadening of its spectrum. Finally, the pulse stabilizes as the transient components decay away. Note that at this stage the flat amplitude background is significantly lower than prior to the pulse formation, a curious result from the modified LLE changing its effective pump field $F'$ in response to the existing pulse in the resonator. The reduced flat amplitude will no longer spontaneously generate patterns, preventing further pulse generation beyond the first pulse. This set of transient steps eventually result in deterministic placement of one broad-band pulse in the resonator, shown in the top panel of Fig. S\ref{Fig:Stability}E.

\vspace{5pt} 
\section*{Disclaimer}
\vspace{-25pt}
\textcolor{black}{This work is a contribution of the U.S. Government and is not subject to copyright. Mention of specific companies or trade names is for scientific communication only, and does not constitute an endorsement by NIST.}

\end{document}